\documentclass[preprint]{aastex}
\usepackage{natbib}
\slugcomment{Accepted for publication by The Astrophysical Journal}
\shorttitle{DETECTION OF EXOMOONS THROUGH POLARIZATION}
\shortauthors{SENGUPTA AND MARLEY }
\received{2016 March 08}
\accepted{2016 April 15}
\begin{document}

\title{DETECTING EXOMOONS AROUND SELF-LUMINOUS GIANT EXOPLANETS THROUGH 
POLARIZATION} 

\author{Sujan Sengupta}
\affil{Indian Institute of Astrophysics, Koramangala 2nd Block,
Bangalore 560 034, India; sujan@iiap.res.in}

\and

\author{Mark S. Marley}
\affil{NASA Ames Research Center, MS-245-3, Moffett Field, CA 94035,  
U.S.A.;
Mark.S.Marley@NASA.gov}

\begin{abstract}
Many of the directly imaged self-luminous gas giant exoplanets have been found to have cloudy atmospheres.
Scattering of the emergent thermal radiation from these planets by the dust grains in their atmospheres should locally give rise to 
significant linear polarization of the emitted radiation. However, the observable disk averaged polarization should
be zero if the planet is spherically symmetric. Rotation-induced oblateness may yield a
net non-zero disk averaged polarization if the planets have sufficiently high spin rotation velocity.
On the other hand, when a large natural satellite or exomoon transits a planet with cloudy 
atmosphere along the line of sight, the asymmetry induced during the transit should
give rise to a net non-zero, time resolved linear polarization signal. 
The peak amplitude of such time dependent polarization may be detectable even for
 slowly rotating exoplanets. Therefore, we suggest that large exomoons around directly imaged self-luminous exoplanets may be
detectable through time resolved imaging polarimetry. Adopting detailed atmospheric models for 
 several values of effective temperature and surface gravity which are
appropriate for self-luminous exoplanets, we present 
the polarization profiles of these objects in the infrared
during transit phase and estimate the peak amplitude of polarization that occurs during the
the inner contacts of the transit ingress/egress phase. The peak polarization is predicted
to range between 0.1 and 0.3 \%  in the infrared.     
    
\end{abstract}


\section{INTRODUCTION}

Natural satellites orbiting all planets except Mercury and Venus are consequential
members of the solar system. The formation of a large variety of
satellites with size ranging from 2600 km to a few kilometers is 
a natural consequence of planet formation. Therefore, it seems
likely
that the abundance of planetary systems discovered around other stars should
also have a significant number of natural satellites or exomoons orbiting the
planets. Indeed both theoretical investigations on the formation and properties as well as
observational searches for exomoons have already been initiated (for a
review see \cite{heller14}.  

    Various methods for detecting exomoons have been suggested and almost all of
them rely upon photometric light curves of the star during the transit of the
planet. For example, baryocentric and photocentric Transit Timing Variation (TTV)
\citep{sartoretti99, szabo06} and Transit Duration Variation (TDV) \citep{kipping09}
rely upon transit photometry. While TTV is more sensitive to wide-orbit moons, TDV is
more sensitive to close-orbit moon. The first and till date the only systematic
program searching for exomoons with Kepler (the Hunt for Exomoons with Kepler program) analyzes  the transit light curves
obtained by Kepler space telescope and attempts to identify the presence of a natural satellite
around the transiting planets\citep {kipping12}.Unfortunately, the effect of an exomoon on
the light curve is so tiny that the expected signature is extremely difficult to identify and therefore, evidence for
an exomoon is still elusive.    

 Apart from  transiting planets,
another class of giant exoplanets that are young, warm and self luminous 
have been discovered by direct imaging \citep[e.g.,][]{Cha04,Mar08,Lag10,Laf10, Mac15}. These planets are typically 10-40 AU away from their parent 
stars. Such a large separation allows them  to be spatially resolved with 
coronagraphic observation as well as differential imaging techniques. 
In the next few years, ground-based telescopes with dedicated high-contrast imaging
instrumentation, such 
as the P1640 coronagraph on Palomar, the Gemini Planet Imager on Gemini South,
SPHERE on the VLT and James Webb Space
Telescope are expected to detect many more such planets \citep{Bei10}. 

Comparisons of synthetic spectra to observed data clearly implies that most of the 
exoplanets directly imaged to date have dusty atmospheres \citep{Mar08,Laf10,Mac15}.
In that respect the atmospheres of these planets are very similar to  those of L dwarfs.
Hence, like the L dwarfs, it is expected that radiation of these planets
as well as those likely to be detected in the future, should be linearly 
polarized in the near-infrared due to scattering by dust grains \citep{marley11}. 
For a homogeneous distribution of scatterers, the net polarization integrated
over the projected disk of a spherically symmetric planet is zero. Horizontal inhomogeneity may yield net non-zero
polarization\citep{deKok11}. On the other hand 
rotation induced oblateness causes asymmetry and gives rise to
significant amount of polarization \citep{sengupta01,sengupta10,marley11}. 
 Observed linear polarization of L dwarfs shows that 
the amount of polarization increases with the increase in spin rotation
velocity \citep{miles13}. This implies that the asymmetry due to  
oblateness plays a dominant role. Apart from  the asymmetry due to  
oblateness, net non-zero disk integrated polarization may arise if the
stellar disk is occulted by a planet. Several authors have 
presented detailed models of such transit polarization of stars of different spectral
types \citep{car05,kostogryz11,wik14,kostogryz15,kostogryz15a}. 
Scattering of light in stellar atmosphere gives rise to linear polarization
 throughout the stellar spectrum. Because of the variation in scattering geometry
 this polarization increases monotonically from  the center of the disk towards the
 stellar limb. 
However, since the polarization of hot stars arises because of electron scattering
and that for cool stars arises because of Rayleigh scattering of atoms and molecules,
the peak amplitude of polarization is extremely small, a few times $10^{-6}$ in the
B-band.
On the other hand, as demonstrated by \cite{sengupta10} and \cite{marley11}, scattering by dust
grains yields high polarization in the infrared. In fact the degree of polarization
detected from  several L dwarfs is as high as 0.1-0.4 \% in the I-band for even
moderately fast rotating L dwarfs.

   In this paper, we suggest that similar to the polarization of a star that arises due
to the transit of a planet,
the self-luminous directly imaged exoplanets should also give rise to
detectable amount of time dependent polarization if the object is eclipsed  by a
sufficiently large natural satellite or exo-moon. Hence, we propose that exomoons
around self-luminous directly imaged planets can be detected through time dependent
image polarimetric observation.     

   In the next section, we briefly describe the atmospheric models adopted. In section~3,
we present the formalism  used to derive the time dependent eclipse polarization by
exomoons of different size and in section~4, we discuss the results. Finally we conclude our
investigation in section~5.

\section{THE PLANETARY ATMOSPHERIC MODELS AND SCATTERING POLARIZATION}

 In order to calculate the intensity and polarization of the planetary 
radiation, we have employed a grid of one-dimensional,
hydrostatic, non-gray, radiative-convective equilibrium
atmosphere models stratified in plane-parallel. These models incorporate about 2200 gas species,
about 1700 solids and liquids for compounds of 83 naturally occurring elements as well as five major
 condensates as opacity sources \citep{ack01,marley02,freed08,Sau08}. 
Silicate and iron clouds computed with sedimentation efficiency $f_{\rm sed}=2$
 are included in the atmospheric model \citep{ack01}. The model atmospheres are
calculated for specified values of effective temperature $T_{\rm eff}$ and surface gravities $g$.

 The atmospheric code computes the temperature-pressure profile, the gas and dust opacity
and the dust scattering asymmetry function
averaged over each atmospheric pressure level. These input data are used in a multiple
scattering polarization code that solves the radiative transfer equations
in vector form. The two Stokes parameter I and Q  are calculated in a locally
plane-parallel medium \citep{sengupta10,marley11}. A combined 
Henyey-Greenstein-Rayleigh phase matrix \citep{Liu06} is used to calculate the
angular distribution of the photons before and after scattering. 
The detailed formalisms as
well as the numerical methods for calculating the angle dependent total
and polarized intensity $I$ and $Q$  are described  in \cite{sengupta09}.
Finally, the angle dependent $I$ and the polarization $P=Q/I$ computed
by the polarization code are integrated over the eclipsed disk of the planet.
The formalism  is described in the next section. We have neglected thermal and
reflected light from the exomoon. 


\section{ECLIPSED DISK INTEGRATED POLARIZATION}

 Similar to the case of a transiting planet, the net polarization during 
eclipse is equal to the fractional circumference blocked by the projection
of the moon over the surface of the planet multiplied by the scattering 
polarization and intensity of the planet at each radial point along
the planetary disk. 

The disk integrated polarization during the eclipse phase is given by \citep{car05,wik14}
\begin{eqnarray}
p_(t) =\frac{1}{F}\int^{r_m(t)+w}_{r_m(t)-w}C(r,t)I(r)P(r)dr,
\end{eqnarray}
where $F$ is the flux of the unobscured planet, $I(r)$ and $P(r)$ are the specific 
intensity and polarization respectively at the
normalized radial co-ordinate $r$ on the disk of the planet,
$w=R_m/R_P$ is the ratio between the radius of the moon ($R_m$) and the radius
of the planet ($R_P$), $C(r,t)$ is the instantaneous path length along the planetary
circumference at $r$ that is eclipsed by the moon and is given as
\begin{eqnarray}
C(r,t)=2\sqrt{[(r-r_m(t)]^2-w^2},
\end{eqnarray}
$r_m(t)$ being the instantaneous position of the center of the
moon and is given by
\begin{eqnarray}
r_m(t)=\left[b^2+4\left\{(1+w)^2-b^2\right\}\left(\frac{t}{\tau}\right)^2
\right]^{1/2},
\end{eqnarray}
where $b=a\cos i/R_p$ is  the impact parameter for circular orbit of radius $a$
and $i$ is the orbital inclination angle of the moon. In the above expression,
$t$ is the time since mid-eclipse and $\tau$ is the eclipse  duration given by
\cite{scharf09}
\begin{eqnarray}
\tau=\frac{P}{\pi}\sin^{-1}\left[\frac{R_p}{a}\left\{\frac{(1+w)^2-b^2}{1-
\cos^2 i}\right\}^{1/2}\right].
\end{eqnarray}
Here, $\mu=\cos\theta$ with $\theta$ being
the angle between the normal to the planetary surface and the line of sight,
$r=\sqrt{1-\mu^2}$, $0\leq r \leq 1$.     
In term of $\mu$, Equation~(1), therefore reduces to      
\begin{eqnarray}
p_(t) =\frac{1}{F}\int^{r_2}_{r_1}2\sqrt{\frac{[(1-\mu^2)^{1/2}-r_m(t)]^2-w^2}{1-\mu^2}}I(\mu)
P(\mu)\mu d\mu,
\end{eqnarray}
where $r_1=\sqrt{1-[r_m(t)+w]^2}$ and $r_2=\sqrt{1-[r_m(t)-w]^2}$.

Similar to the orbital period
of Ganymede around Jupiter, we have fixed the orbital period of the exomoon
at 7 days and have calculated the orbital distance by using Kepler's law. The radius
of the planet is fixed at 1$R_J$ where $R_J$ is the radius of Jupiter. The mass of the
planet is calculated from the surface gravity of the planet. It is worth mentioning that
this is just a representative case. The orbital period does not alter the
amount of polarization originated due to eclipse but it determines the interval between
the two successive amplitudes of polarization that occur at the inner contact points of
the transit ingress/egress phase and provides some qualitative guidance to the
order of magnitude duration of the polarization event.
The polarization profile is calculated for two values of
the inclination angle for the satellite,  $90^o$ and $88^o$. Similar to the case
of transiting planets, eclipse cannot occur if the inclination angle
$i\leq \cos^{-1}\left(\frac{R_p+R_m}{a}\right)$.

\section{RESULTS AND DISCUSSION}

For transit polarization models of stars, the local angle dependent polarization
arises by scattering of light with atoms and molecules and the maximum amount of 
polarization that occurs at the stellar limb is usually
very small. Scattering polarization at different angular or radial points of a
solar type star varies from a few times of
 $10^{-4}$ near the
limb to a few times of $10^{-6}$ near the center for wavelengths in the range of
4000-5500 $\AA$ \citep{fluri99}. At longer wavelengths the polarization is much smaller
than $10^{-6}$.
The polarization is of course zero at the center
($\mu=1$) and maximum at the limb ($\mu=0$).
The contribution of a scattering opacity to the total opacity in the atmosphere 
is responsible for the center-to-limb variation in the polarization across any stellar disk
\citep{harrington69}. In hot stars (spectral types O, B or A), electron scattering gives rise to such opacity while
 Rayleigh scattering by atomic and molecular hydrogen and atomic 
helium  is the main source of scattering opacity in cool stars. The polarization at the limb due
 to electron scattering is as high as 0.1 \citep{chandra} while the polarization for
 wavelengths of solar resonant lines such as Ca or Sr, is about 0.16 \citep{bianda99}.
 This is about three orders of magnitude higher than the polarization at the solar
limb observed in the solar continuum \citep{fluri99}. On the other hand, 
if the planetary radiation is polarized by dust 
scattering then the amount of polarization at the limb for the far optical and infrared
wavelength should be higher than that arises due to electron scattering at near
optical wavlength. As realized  by
\cite{car05}, formation of dust grains in the cloudy atmosphere of L brown dwarfs would
provide an additional scattering opacity to the gas opacity which should yield into
large values for limb polarization. The self-luminous giant exoplanets are also
 expected to have cloudy atmosphere and linear polarization as large as 1-2\%  in the
 infrared is already predicted by \cite{marley11}.    

  In order to check the validity of the formalism adopted and the correctness of the numerical
calculations, we reproduced the transit polarization profiles presented by \cite{car05}.
These authors used the analytical expressions for the limb darkening law suggested by 
\cite{claret00} and for the solar continuum  polarization suggested by \cite{fluri99}. 
\cite{car05} also estimated the maximum transit polarization of a T-dwarf of radius
$0.2\,\rm R_{\odot}$ occulted by an earth size exoplanet. For  $R_m / R_P=0.046$, 
\cite{car05} found  the peak amplitude of transit polarization at wavelength 4600 $\AA$ is
$1.6\times 10^{-4}$ or 0.016\%.. 
As presented in figure~1, for the
same value of $R_m/R_P$ and for a central eclipse (i=$90^o$), we find the peak polarization
to be 0.02\%  at the B-band, in good
agreement with the result of \cite{car05}. Figure~1 presents the eclipse polarization
profiles at B-, I-, J- and H-bands of a self-luminous cloudy exoplanet of radius 1$R_J$.
At B-band, Rayleigh scattering by molecules dominates over dust scattering. As the wavelength
increases, dust scattering becomes more important in determining the amount of polarization.
Figure~1 shows that the ecplise polarization profiles for B- and I-bands are almost the same
but the peak polarization increases by about six times at J-band. That the polarization at
J-band is much higher than the polarization at I-band is already demonstrated by
\cite{marley11} for a variety of atmospheric and cloud models. The polarization 
however, decreases at H-band as the wavelength of the radiation increases 
farther.  
The wavelength dependency of the polarization is governed by the adopted cloud model,
 e.g., the size distribution, the number density of the dust grains as well as the 
location of the cloud base and deck.

     The disk integrated polarization of a spherical exoplanet eclipsed by an
exomoon is presented in figure~2 for two different values of the orbital inclination
angles and three different sizes of the exomoon. The disk integrated polarization of a 
unobscured oblate planet is also presented for comparison. Disk integrated
polarization of a rotation-induced non-spherical and eclipsed planet could be
complicated owning to the fact that the net polarization depends on the
angle between the spin axis of the planet and the orbital plane of the exomoon. Since
it is not possible to determine the spin axis orientation of the exoplanet, 
we restrict our  polarization models for  slow
rotating eclipsing planets for which the departure from sphericity is too 
small to yield net non-zero polarization when it is not in eclipsing phase. The
projected spin angular velocity of a few  self-luminous exoplanets has already been
derived from observation \citep{Mar10,Snell14} which can provide the minimum value
 of the oblateness of the
object. Disk integrated polarization for oblate self-luminous giant
exoplanets has been presented by \cite{marley11} for various values of the spin period,
surface gravity and effective temperature. We have neglected any inhomogeneous
distribution of scatters. Our atmospheric models consider vertically inhomogeneous but
horizontally homogeneous cloud distributions.

   As shown in figure~2, the qualitative feature of the eclipse polarization profile
is the same to that of transit polarization presented  by \cite{car05,kostogryz11} and
\cite{kostogryz15}. The double peaked polarization profile arises because of the fact that
the maximum polarization occurs near the inner contacts of ingress/egress phases. For
central eclipse i.e., when the inclination angle $i=90^o$, the projected position of the
the center of the moon during mid eclipse is at the center of the planet causing a symmetry to the projected
stellar disk. Hence
the disk integrated polarization for central eclipse is zero during mid eclipse. The
polarization increases as the moon moves from the center ($t=0$)to the limb (t=$\pm\tau/2$)
of the planetary disk as it induces asymmetry to the planetary disk. However,
when the eclipse is off center, i.e., $i < 90^o$, the polarization is non-zero 
during the whole eclipse epoch. The net polarization at the mid eclipse increases
with the decrease in the orbital inclination angle of the moon. However, the peak
polarization at the limb remains the same irrespective of the inclination angle.
Therefore, the peak polarization
is independent on the orbit inclination but depends on the radii ratio of
 satellite to planet.  The transit duration
depends on the size of both the moon and the planets, on the orbital distance of the moon 
from the planet and on the inclination angle. Therefore, for a given orbital
distance, the peak polarization occurs at different time for different size ratio and
inclination angle. Figure~2 shows that the peak polarization at the inner contacts of
the ingress/egress
phase increases linearly  with the increase in the size of the exomoon. 
At J-band the maximum degree of linear polarization for $R_m/R_p=0.046$
 is 0.11\% while that for $R_m/R_p$=0.07 and 0.1 are 0.16\% and  0.21\%.
 respectively.  In the absence of an eclipsing exomoon, 0.1\%  of linear
polarization can  arise only if the spin-rotation period of the planet with surface 
gravity $g=30 {\rm ms^{-2}}$ is less than 6.6 hours while it needs
a spin rotation period less than 6.1 hours to yield net non-zero disk integrated polarization
of degree 0.2 \%. For higher surface gravity, an even shorter period 
is required in order to yield
sufficient oblateness to the planet. Therefore, the time dependent eclipse polarization can be distinguished from
the disk integrated  polarization of an oblate planet if the spin rotation
period of the planet is higher than 10-15 hours so that the asymmetry due to spin-rotation
is negligible.

  \cite{marley11} and \cite{sengupta10} have demonstrated that the degree of
polarization of cloudy self-luminous exoplanets and L dwarfs depends on the 
effective temperature and  the surface gravity of the objects. In figure~3, 
we present the eclipse polarization profiles for exoplanets with different effective
temperature and surface gravity. At the lowest surface gravity $g=30 {\rm ms^{-2}}$ among the
three cases we consider, the highest
polarization is obtained at J-band when $T_{eff}=800$ K. If the temperature increases
further, the cloud base shifts upward yielding a smaller column of   dust grains in the observed atmosphere
and hence the polarization decreases with the increase in effective temperature.
 A balance between the downward transport by sedimentation and upward turbulent diffusion of
condensates and gas determines the scattering opacity and hence the polarization for different
effective temperature and surface gravity.  
For $g=56 {\rm m s^{-2}}$, the polarization is maximum at $T_{eff}=1000$K and it
decreases both with the increase and decrease in effective temperature. In fact 
 figure~3 does not show
much difference in polarization for $T_{eff}$=1200 and 1000K.  
 On the other hand, at the highest surface gravity considered here,
 $g=100{\rm ms^{-2}}$, the polarization
is greatest at $T_{eff}=1200\,\rm K$ but decreases at lower effective temperatures. Therefore, a combination of the surface gravity and effective
temperature along with the size of the exomoon as compared to the exoplanet determine
the amount of polarization caused by the eclipse. However, the difference in the amount
of polarization among these cases is likely too small to be differentiated observationally. Moreover, the
 effect of surface gravity and effective temperature can be realized even without the
eclipsing effect. For example, a sufficiently fast rotator that may yield detectable
amount of net non-zero polarization would show the same properties of the polarization 
profile. 
 Finally, as seen from figure~3, if $T_{eff}$ is as low as 600 K then for any surface gravity,
the polarization reduces substantially owing to less amount of dust grains in the 
atmosphere probed in the near-infrared in these models.
It is worth mentioning that the orbital separation or the inclination angle
does not alter the peak amplitude of polarization but alters the overall polarization
profile.

\section{CONCLUSIONS}

   We suggest that time resolved  imaging polarimetry may be a potential technique
to detect
large exomoon around directly imaged self luminous exoplanets. The cloudy atmosphere of
such planets ensure high polarization due to scattering by dust grains. However, the 
disk-integrated net polarization would be zero if the planet is
spherically symmetric. This symmetry is broken if either the planet
loses sphericity due to fast spin rotation, has horizontally inhomogeneous cloud
structures or if a sufficiently large satellite eclipses
the planetary surface. Asymmetery by a combination of more than one of the above
causes is also possible. Assuming a spherical exoplanet eclipsed by an exomoon,
we estimated the peak polarization at the inner contact points of ingress/egress phase and
presented the eclipse polarization  profile. Our investigation implies that detectable 
amount of polarization may arise in J-band if the planet is eclipsed by a large exomoon.
For a central eclipse, the detectable amount of polarization would arise only at the
inner contacts of ingress/egress phase while for off central eclipse, polarization
may be detectable during the entire eclipsing phase. Unlike the constant polarization
that may be detected for fast rotating exoplanet, no polarization will be detected when
the planet is out of the eclipse phase.

 According to our estimation, an image polarimeter with sensitivity ranging from 0.3 to
0.01 \% may detect the presence of exomoon around the self-luminous exoplanets that
are directly imaged. Future high-contrast imaging instruments on thirty-meter class
telescopes may plausibly provide both the time resolution and polarimetric sensitivity to detect such moons.

\section{Acknowledgements} We thank the reviewer for a critical reading of the
manuscript and for providing several useful suggestions.

\clearpage
\begin{figure}
\includegraphics[angle=0.0,scale=0.8]{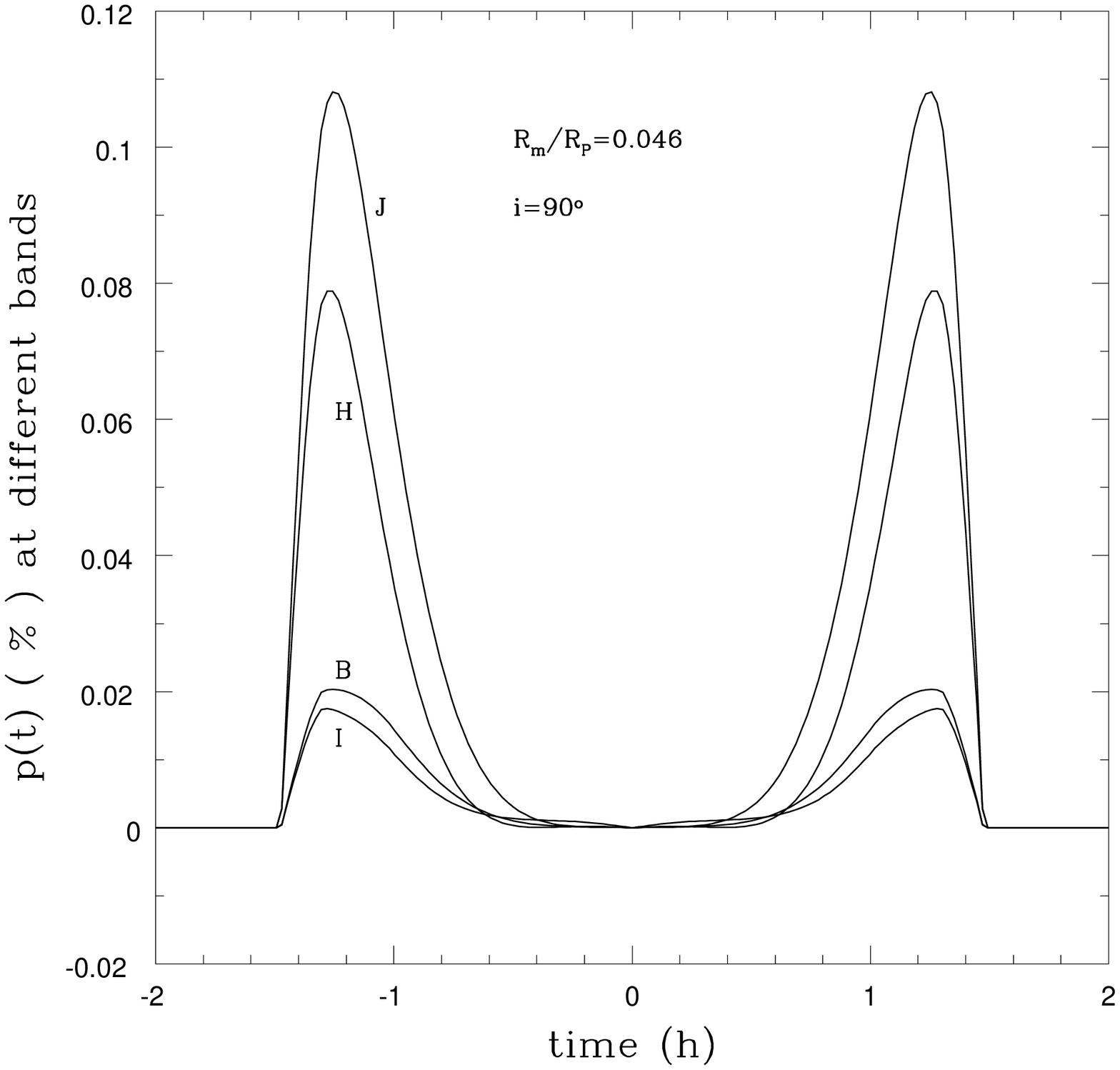}
\caption{Disk integrated linear polarization at different wavelength bands of a
self-luminous spherical exoplanet partially eclipsed by a moon of radius 
0.046 times the radius of the planet.
\label{fig1}}
\end{figure}

\begin{figure}
\includegraphics[angle=0.0,scale=0.8]{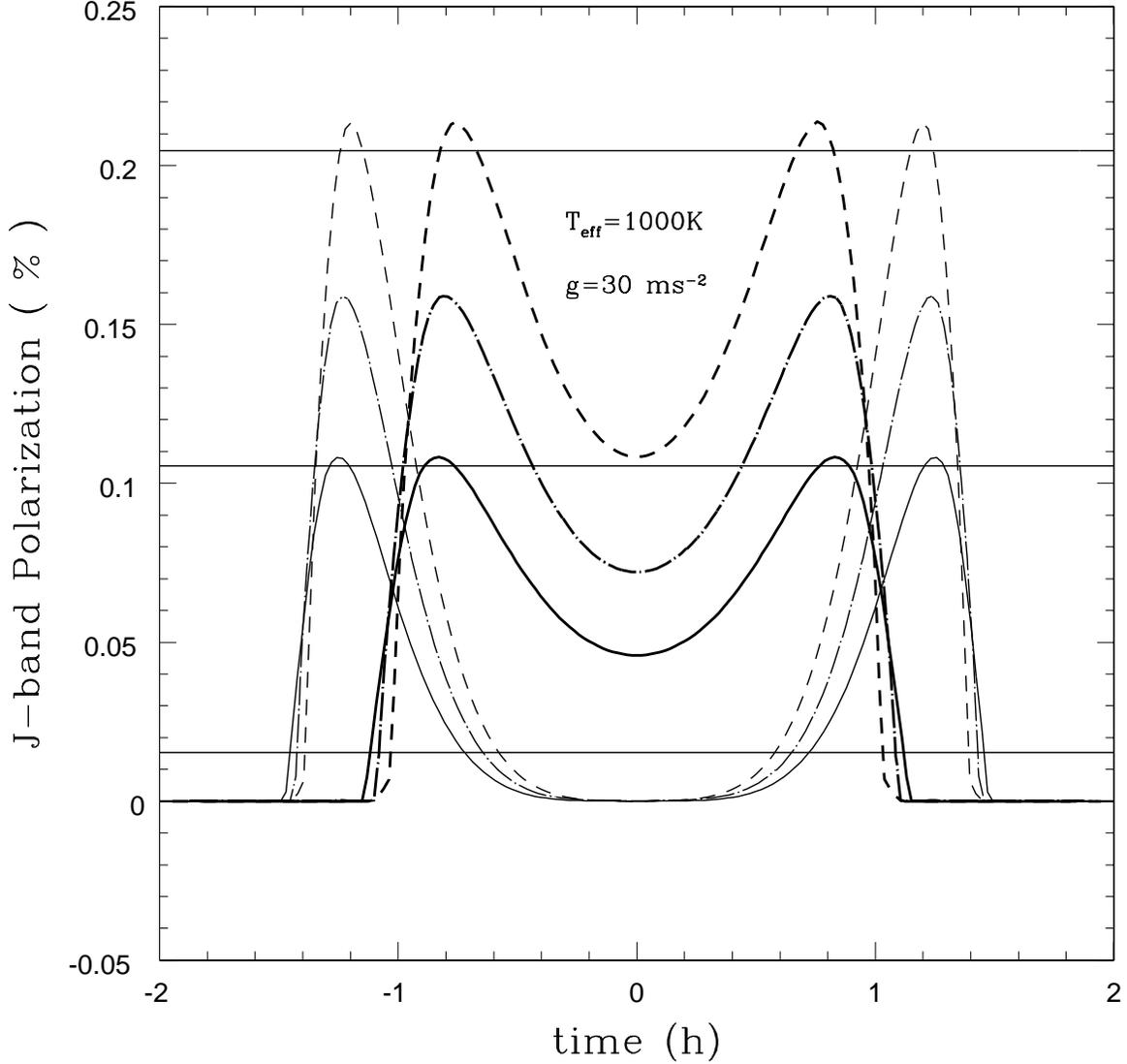}
\caption{J-band disk integrated polarization of self-luminous, 
spherical exoplanets partially eclipsed by an exomoon. Thin and thick 
lines represent the percentage polarization for 
an exomoon transit with an orbital inclination angle of $i=90^o$ and $88^o$ 
respectively.  Solid lines, dashed-dot lines and dashed lines represents eclipse
polarization profile with $R_m/R_P$=0.046, 0.07 and 0.1 repectively.
The orbital period of the exomoon is set to 7 days for all cases.
From top to bottom, the horizontal lines represent 
linear polarization integrated over the disk of a rotation-induced 
oblate exoplanet (with no eclipse) with spin period 6.1, 6.6, and 15 hours 
respectively.  
\label{fig2}}
\end{figure}

\begin{figure}
\includegraphics[angle=0.0,scale=0.8]{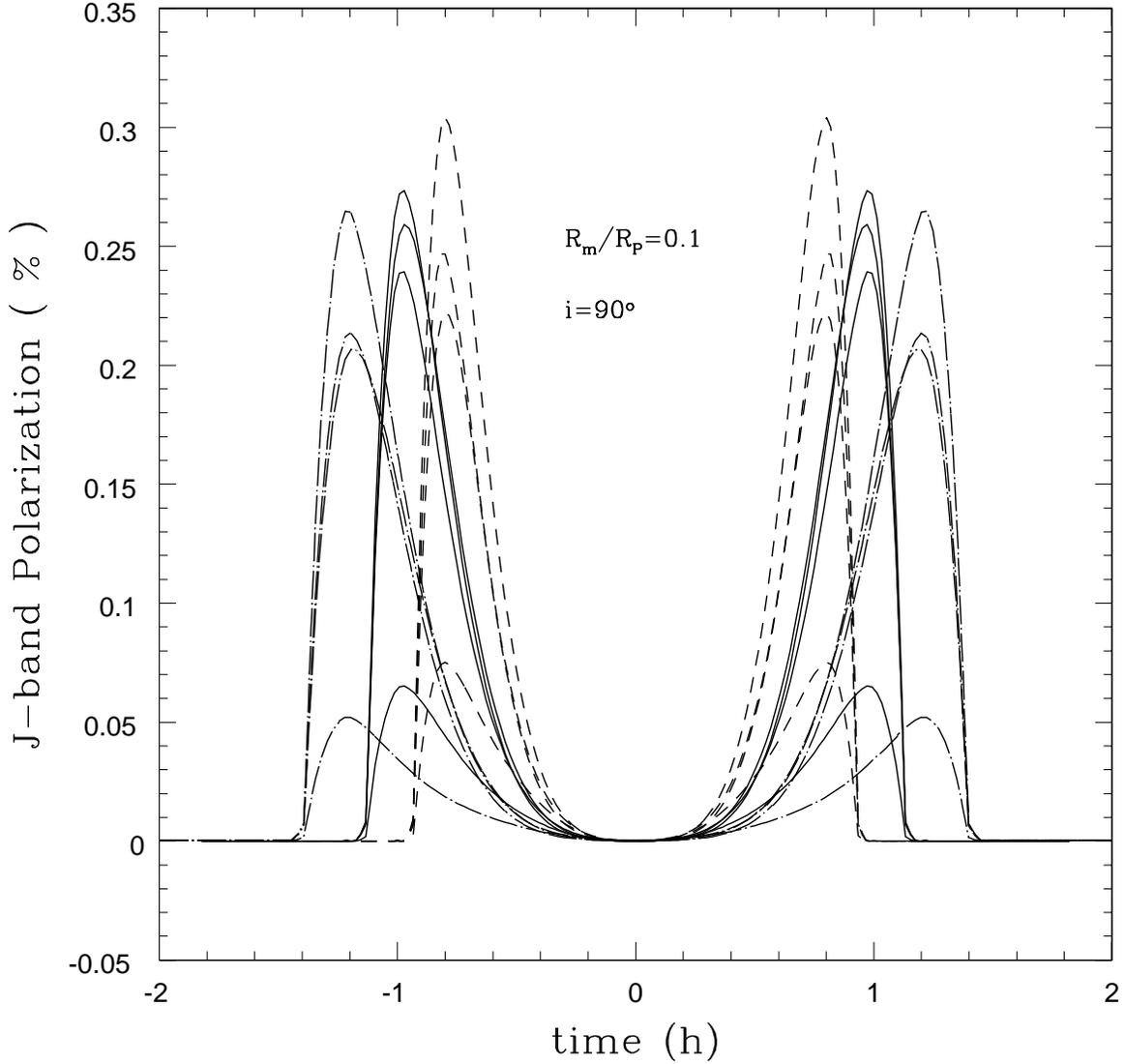}
\caption{ J-band polarization of exoplanets with different effective temperature
and surface gravity. Solid lines represent exoplanetary models with $g=56 
{\rm ms^{-2}}$; from top to bottom the solid lines represents models with
$T_{eff}=$1000, 1200, 800 and 600 K. Dashed lines represent the
polarization with $g=100 {\rm ms^{-2}}$ and from top to bottom  dashed lines
represent models with $T_{eff}$=1200, 1000, 800 and 600 K. Similarly, 
dot-dashed line represents model with $g=30 {\rm  ms^{-2}}$ and from top to
bottom, the dot-dashed  lines represent exoplanet models with 
$T_{eff}$=800, 1000, 1200 and 600 K.
\label{fig3}}
\end{figure}

\clearpage
\end{document}